\documentclass[conference]{IEEEtran}
\IEEEoverridecommandlockouts
\usepackage{cite}
\usepackage{amsmath,amssymb,amsfonts}
\usepackage{algorithmic}
\usepackage{graphicx}
\usepackage{textcomp}
\usepackage{xcolor}
\usepackage{booktabs} 
\def\BibTeX{{\rm B\kern-.05em{\sc i\kern-.025em b}\kern-.08em
    T\kern-.1667em\lower.7ex\hbox{E}\kern-.125emX}}
\begin{document}

\title{Layout-Aware OCR in Black Digital Archives: An Unsupervised Evaluation Approach

}

\author{\IEEEauthorblockN{Fitsum Sileshi Beyene}
\IEEEauthorblockA{\textit{Department of Computer Science and Engineering} \\
\textit{The Pennsylvania State University}\\
University Park, PA, USA \\
fsb5110@psu.edu}
\and
\IEEEauthorblockN{Christopher L. Dancy}
\IEEEauthorblockA{
\textit{Department of Industrial and Manufacturing Engineering} \\
\textit{The Pennsylvania State University}\\
University Park, PA, USA \\
cdancy@psu.edu}

}

\maketitle

\IEEEpubid{\begin{minipage}{\textwidth} \vspace{5\baselineskip} \centering\footnotesize
This version of the manuscript is the original author-submitted version to IEEE International Symposium on Technology and Society (ISTAS 2025). Please cite final version when available on IEEE Xplore.
\end{minipage}}
\IEEEpubidadjcol

\begin{abstract}
Despite their cultural and historical significance, Black digital archives continue to be a structurally underrepresented area in AI research and infrastructure. This is especially evident in efforts to digitize historical Black newspapers, where inconsistent typography, visual degradation, and limited annotated layout data hinder accurate transcription, despite the availability of various systems that claim to handle optical character recognition (OCR) well. In this short paper, we present a layout-aware OCR pipeline tailored for Black newspaper archives and introduce an unsupervised evaluation framework suited to low-resource archival contexts. Our approach integrates synthetic layout generation, model pretraining on augmented data, and a fusion of state-of-the-art You Only Look Once (YOLO) detectors. We used three annotation-free evaluation metrics, the Semantic Coherence Score (SCS), Region Entropy (RE), and Textual Redundancy Score (TRS), which quantify linguistic fluency, informational diversity, and redundancy across OCR regions. Our evaluation on a 400-page dataset from ten Black newspaper titles demonstrates that layout-aware OCR improves structural diversity and reduces redundancy compared to full-page baselines, with modest trade-offs in coherence. Our results highlight the importance of respecting cultural layout logic in AI-driven document understanding and lay the foundation for future community-driven and ethically grounded archival AI systems.
\end{abstract}


\section{Introduction}
The digitization of historical newspapers has enabled the preservation and access to culturally significant materials, especially in history, journalism studies, and public memory [1], [2]. However, many community-specific archives, such as Black newspapers remain under-digitized, both in terms of scanning coverage and downstream computational accessibility [3], [4]. These publications document local civil rights efforts, social commentary, and political advocacy, but were historically printed under resource-constrained conditions and with highly variable layouts. At times, these layouts also constituted a design language, which functioned in concert with the words written to portray (at times \textit{fugitive} [31]) meaning. Consequently, their visual irregularity, typographic inconsistency, and physical degradation challenge conventional document analysis pipelines [5],[6],[7], even those meant to process historical archives.

Optical character recognition (OCR) systems depend critically on accurate layout analysis to detect, group, and sequence text regions on a page. Without it, the system may read text out of order, duplicate it, or fragment it, particularly in low-resource and multicolumn documents with overlapping typographic zones [8]. However, many OCR systems, including open source engines such as Tesseract and EasyOCR, lack optimization for historical materials and often fail when processing atypical layout geometries or degraded scans [9]. The problem is exacerbated in archives that lack layout annotations, making it impossible to train or benchmark supervised layout detection models [10].

Although deep learning-based object detectors, such as the YOLO family of models, have achieved strong performance on layout segmentation in modern documents [11], [12], they rely on large, domain-specific datasets. Public benchmarks such as PubLayNet [13], DocBank [10], and DocLayNet [14] consist primarily of scientific PDFs and office documents. Models trained on these datasets do not generalize well to materials featuring irregular historical typography. Recent efforts to adapt layout models to archival settings have shown some promise [15], but most target European newspapers or assume access to high-quality ground truth, which is rarely available in Black digital archives [3], [4].

This work presents a layout-aware OCR pipeline developed in response to the visual and structural challenges present in historical Black newspaper archives published between 1827 and 1859. We curated a 400 page benchmark dataset derived from 26 digitized multi-page newspapers, including titles such as Freedom’s Journal, The Colored American, Mirror of Liberty, The North Star, Frederick Douglass’ Paper, and The Weekly Anglo-African. These materials display a wide range of layout styles and degradation artifacts, providing a representative testbed for evaluating document understanding in data-scarce archival contexts.

A key limitation in working with these archives is the absence of a gold-standard layout or OCR ground truth. To address this, we used an unsupervised evaluation framework composed of three metrics: the Semantic Coherence Score (SCS), which measures the proportion of recognized dictionary words per region; the Region Entropy Divergence (RED), which captures the diversity of n-gram patterns across segments; and the Textual Redundancy Score (TRS), which penalizes repeated transcriptions across overlapping boxes. These metrics offer a scalable and annotation-free method to compare layout hypotheses in OCR pipelines, particularly in data-scarce archival domains [16], [17], [18], [19].

This study forms the basis for a forthcoming interdisciplinary workshop focused on ethical and liberatory AI development for Black digital archives. The pipeline and evaluation framework presented here are intended not only to support technical advances but also to enable critical engagement with archival AI systems by scholars in black studies, digital humanities, and socio-technical research. By foregrounding layout as both a computational and cultural concern, we aim to create space for future co-designed systems that respect the specific structural logic of media from historically marginalized (especially Black) communities.

\section{Related Work}

Researchers have increasingly used deep learning for document layout analysis (DLA), especially with object detectors like YOLOv8 and YOLOv10, achieving strong results on modern, digitally-born documents [11], [20]. These models are typically trained on large-scale public benchmarks like PubLayNet, DocBank, and DocLayNet, which feature clean, regular layouts [10], [13], [14]. However, these tools often fail when applied to historical documents that include visual degradation, non-standard layouts, and uncommon typographies. Zhao et al. [11] and Gao and Li [21] propose extensions of YOLO specifically designed for layout analysis, but the models are primarily validated on modern or synthetic layouts.

To address challenges in low-resource historical archives, researchers have explored weak supervision and synthetic augmentation. Saha et al. [22] use deep learning with transfer learning and weak supervision for graphical object detection in documents, while Shehzadi et al. [23] experiment with hybrid CNNs and heuristics for regional zoning in European newspaper archives. Araújo [24] applies layout analysis to German-Brazilian materials with Gothic typography. However, relatively little prior work has modeled fine-grained categories like articles, subheadings or advertisements, which are common in newspapers but underrepresented in public benchmarks. 

Evaluating OCR systems without gold-standard transcriptions remains a persistent challenge, particularly in under-annotated archives [25]. Lexicon-based metrics, such as the proportion of recognized words matching entries in a reference dictionary, have been used as unsupervised indicators for assessing recognition quality in handwritten text recognition [26]. Additionally, entropy-based measures have been utilized to evaluate the diversity and distributional characteristics of model predictions in document analysis tasks [26][27].

Layout communicates cultural and rhetorical meaning, especially in publications from historically marginalized communities [29]. Saha et al. [22] demonstrate that modern segmentation models often fail on structurally complex historical documents, necessitating the development of specialized approaches. For example, Black press editors used layout in \textit{The North Star} and \textit{The Weekly Anglo-African} not just for organizing content, but to emphasize rhetoric, build social cohesion, and assert political identity. Barlindhaug [28] emphasizes that the form and materiality of historic documents should be respected throughout digitization and computational analysis. Participatory and justice-oriented approaches in archival AI design, as advocated by Morrison [31], call for the direct involvement of historians, curators, and affected communities in both the interpretation and deployment of computational models.

\section{Methodology}

The objective of our methodology is two fold: first, to develop a layout-aware OCR pipeline specifically adapted to Black historical newspaper archives; and second, to use an unsupervised evaluation framework for layout quality assessment in the absence of ground-truth transcriptions.

\subsection{Manual Annotation and Class-Aware Augmentation}\label{sec:annotation}
Our dataset consists of 400 high-resolution grayscale pages from 26 digitized issues across ten major African American newspapers published between 1827 and 1859 (see Table 1 for details). Eighty-five pages were selected for manual annotation using \texttt{Label Studio}, with regions labeled article, headline, subheading, or advertisement. 

\begin{table}[htbp]
\caption{Black Historical Newspapers dataset}
\begin{center}
\begin{tabular}{|l|c|c|}
\hline
\textbf{Newspaper Title} & \textbf{First Page Year} & \textbf{Pages Used} \\
\hline
\textit{Freedom’s Journal} & 1827 & 208 \\
\textit{Impartial Citizen Vol 1} & 1851 & 82 \\
\textit{The Colored American} & 1837 & 31 \\
\textit{Mirror of Liberty} & 1838 & 26 \\
\textit{Rams Horn} & 1847 & 11 \\
\textit{The North Star} & 1847--1848 & 16 \\
\textit{Frederick Douglass' Paper} & 1855--1859 & 16 \\
\textit{The Self Elevator} & 1853 & 4 \\
\textit{Weekly Anglo-African} & 1859 & 4 \\
\textit{The  Aliened American} & 1853 & 4 \\
\hline
\textit{Total} & 1827-1859 & 400 \\

\hline
\end{tabular}
\label{tab:newspapers}
\end{center}
\end{table}

To address class imbalance among region types, we applied a class-aware augmentation using the \texttt{Albumentations} library. For each underrepresented class, we generated augmented variants of each element by applying: random brightness/contrast adjustment ($p = 0.5$), small-angle rotation (within $\pm 2^\circ$), elastic deformation  ($p = 0.3$), and Gaussian blur ($p = 0.2$)). This produced three augmented samples per original element, expanding the diversity of rare classes for downstream model training. 
    
Using the annotated regions, we extracted bounding box elements and modeled their geometric properties via Gaussian kernel density estimation (KDE). This enabled the generation and evaluation of 1,500 synthetic newspaper pages with respective pseudo-annotation, each constructed with a centered headline, 2–7 columns containing articles, optional subheadings, and advertisements. 

\subsection{Model Training and Evaluation}

\subsubsection{Pretraining on Synthetic Data}

We pre-trained a YOLOv10m model on the synthetic dataset of 1{,}500 page images described in Section~\ref{sec:annotation}. Training was conducted using the Ultralytics implementation of YOLO on an  NVIDIA® GeForce RTX™ 4090. The model was trained for 150 epochs using stochastic gradient descent with a cosine learning rate schedule, an initial learning rate of 0.001, batch size of 8, and input resolution of 1280$\times$1280. This pre-trained model is denoted \textbf{YOLOv10-P}.

\subsubsection{Fine-Tuning with Real Data}

The 85 newspapers with manual annotations (see Section~\ref{sec:annotation}) were randomly split into 80\% training (68 images) and 20\% validation (17 images). We fine-tuned three model variants using identical training hyperparameters: YOLOv8m, YOLOv10m, and YOLOv10-P.

Each model was trained for 250 epochs with identical optimizer and batch settings.

\subsubsection{Model Fusion and OCR Integration}
To integrate predictions from multiple models (YOLOv8, YOLOv10, YOLOv10-P), we implemented a fusion module that:

\begin{enumerate}
  \item Groups boxes with Intersection-over-Union (IoU) $\geq$ 0.7 and identical class labels
  \item Computes a confidence-weighted average of bounding box coordinates
  \item Suppresses near-duplicate boxes
\end{enumerate}

This yielded a set of layout predictions per image that benefits from consensus across model variants.

Each fused region was then processed using a custom OCR pipeline based on \texttt{Pytesseract}. Regions were first preprocessed with denoising, local contrast enhancement (CLAHE), and adaptive thresholding to improve legibility in degraded scans. For tall or noisy regions, we used a sliding-window OCR strategy: each region was divided into overlapping horizontal strips, text was then extracted from each window, and redundant lines were filtered before concatenation. The resulting transcriptions were stored in a modular format:

\begin{verbatim}
{
  "bbox": [...],
  "label": "article",
  "ocr_text": "...",
  "confidence": ...
}
\end{verbatim}

\subsection{Unsupervised Evaluation Metrics}

To assess layout quality in the absence of ground-truth, we used three unsupervised metrics: Semantic Coherence Score (SCS), Region Entropy Divergence (RED), and Textual Redundancy Score (TRS).

\subsubsection{Semantic Coherence Score (SCS)}

SCS is the mean fraction of valid dictionary words across OCR regions:

\begin{equation}
\text{SCS}(T) = \frac{1}{n} \sum_{i=1}^n \frac{|\{ w \in T_i \mid w \in \mathcal{V} \}|}{|T_i|}
\end{equation}

Here, $T_i$ is the text of region $i$, and $\mathcal{V}$ is a reference English word list.

\subsubsection{Region Entropy Divergence (RED)}

RED captures diversity of $n$-gram patterns using Shannon entropy:

\begin{equation}
\text{RED}(T) = - \sum_{g \in \mathcal{G}} P(g) \log_2 P(g)
\end{equation}

where $P(g)$ is the probability of an $n$-gram $g$ in the set of observed $n$-grams $\mathcal{G}$. Higher RED values indicate structurally distinct segments.

\subsubsection{Textual Redundancy Score (TRS)}

TRS penalizes repeated text content across overlapping regions:

\begin{equation}
\text{TRS}(T) = \frac{|\{ (i, j) \mid i \ne j \land T_i = T_j \}|}{n(n - 1)}
\end{equation}

Lower TRS is better, indicating non-redundant region transcriptions.

 \section{Results and Evaluation}

This section presents an evaluation of our document layout parsing and OCR pipeline. We observed detection performance across three fine-tuned object detection models—YOLOv8-m, YOLOv10-m, and YOLOv10-P —followed by downstream OCR evaluation using unsupervised textual quality metrics.

\subsection{Layout Detection Performance}

We evaluated two state-of-the-art object detection models—YOLOv8-m and YOLOv10-m—on a benchmark of historical Black newspaper pages with annotated layout regions.

\textbf{YOLOv8-m} (fine-tuned) achieved the highest overall layout detection accuracy, with a mean Average Precision at IoU 0.5 (\textbf{mAP@0.5}) of 0.736 and mAP@0.5:0.95 of 0.462. Precision and recall were robust for articles and headlines, with moderate performance on subheadings and advertisements.

\textbf{YOLOv10-m} (fine-tuned) performed comparably, with overall mAP@0.5 of 0.716 and mAP@0.5:0.95 of 0.438. Article and headline detection were similar to YOLOv8-m, but subheading and advertisement accuracy was slightly lower:

\begin{figure}[!ht]
    \centering
    \includegraphics[width=0.9\linewidth]{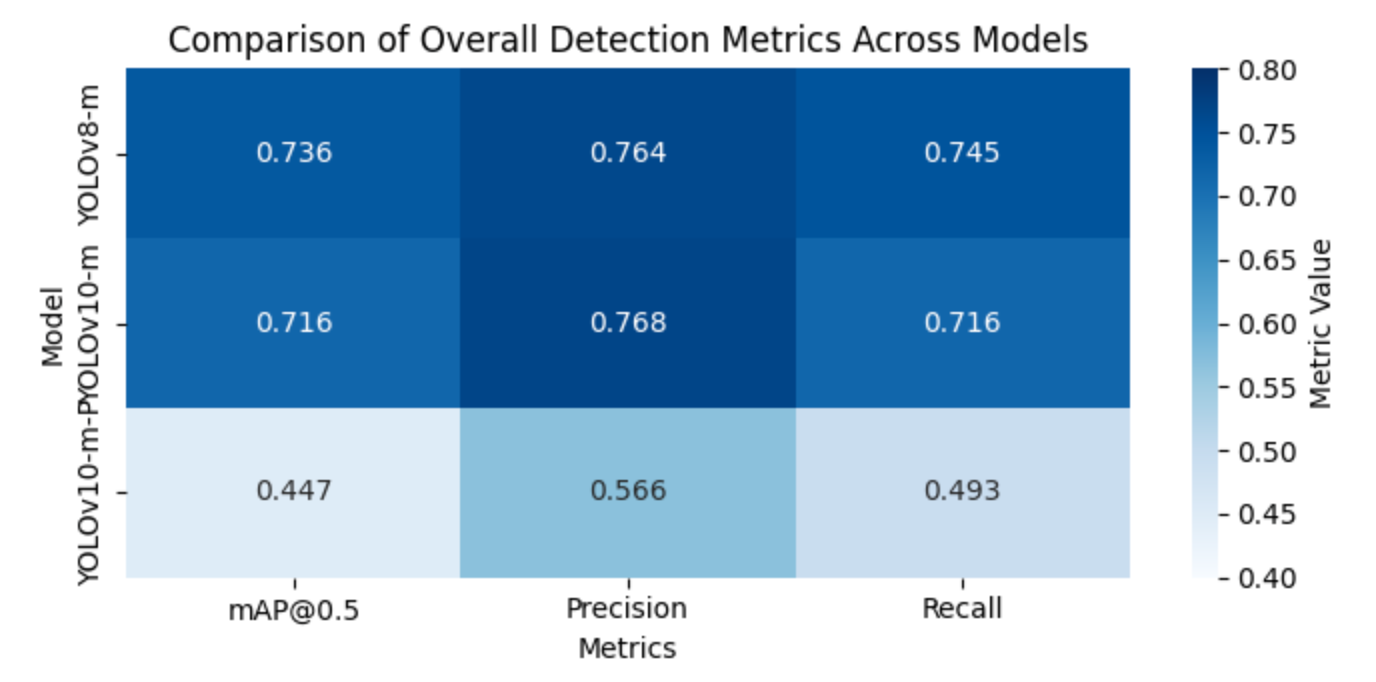 }
    \caption{Comparison of overall detection metrics (mAP@0.5, Precision, Recall) for YOLOv8-m, YOLOv10-m, and YOLOv10-m pretrained on synthetic data. }
    \label{fig:model-metric-heatmap}
\end{figure}

\subsection{Unsupervised OCR Evaluation}

To quantify the downstream utility of layout parsing, we performed OCR over detected regions and compared two transcription pipelines:

\begin{itemize}
    \item \textbf{Layout-based OCR}: Applied to each detected region.
    \item \textbf{Whole-image OCR}: Applied directly to the full page.
\end{itemize}

\begin{table}[h]
\centering
\caption{Average Unsupervised OCR Evaluation Metrics for Each Pipeline}
\label{tab:ocr_results}
\begin{tabular}{lccc}
\toprule
\textbf{Model} & \textbf{SCS} $\uparrow$ & \textbf{RED} $\uparrow$ & \textbf{TRS} $\downarrow$ \\
\midrule
FullPage   & \textbf{0.555} & 11.146 & \textbf{0.000} \\
Fusion     & 0.491 & \textbf{11.843} & 0.001 \\
YOLOv10    & 0.496 & 11.532 & 0.006 \\
YOLOv10p   & 0.505 & 11.559 & 0.005 \\
YOLOv8     & 0.496 & 11.697 & \textbf{0.000} \\
\bottomrule
\end{tabular}
\end{table}

\subsubsection{Semantic Coherence and Diversity}

As shown in Table~\ref{tab:ocr_results}, the Full Page pipeline yields the highest SCS, with a semantic coherence advantage of about 10\% compared to layout-based models. In contrast, all layout-based models outperform Full Page in structural diversity, with the best-performing model (Fusion) achieving a 6\% higher RED and the YOLO models showing 3--5\% improvements.

\subsubsection{Textual Redundancy}

 As shown in Table~\ref{tab:ocr_results}, both the Full Page and YOLOv8 pipelines achieve a score of 0.000, indicating no redundancy in the OCR outputs. The Fusion and YOLO-based layout models also maintain extremely low TRS values, with Fusion scoring 0.001, YOLOv10p at 0.005, and YOLOv10 at 0.006. 

\begin{figure}[ht]
    \centering
    \begin{minipage}{0.48\linewidth}
        \centering
        \includegraphics[width=\linewidth]{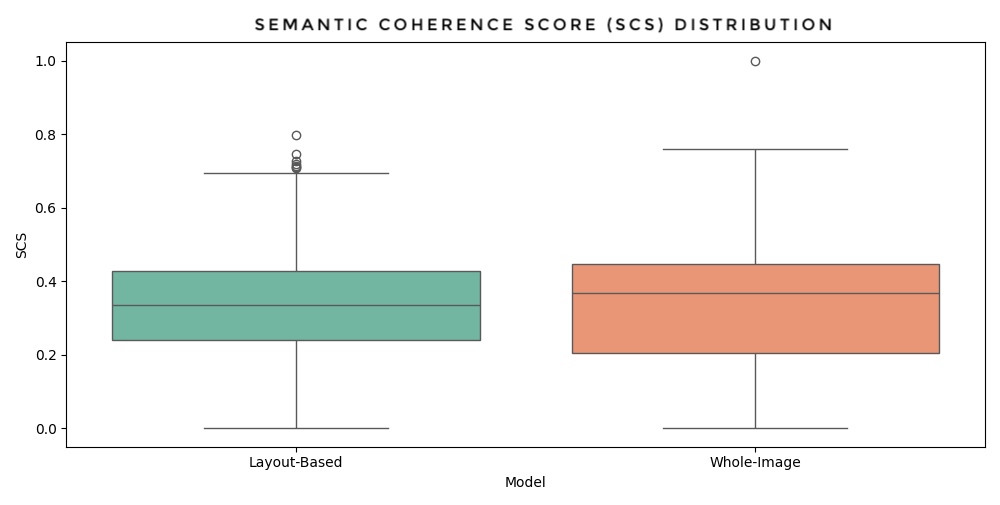}
        \caption{Distribution of Semantic Coherence Score (SCS)}
        \label{fig:scs_boxplot}
    \end{minipage}\hfill
    \begin{minipage}{0.48\linewidth}
        \centering
        \includegraphics[width=\linewidth]{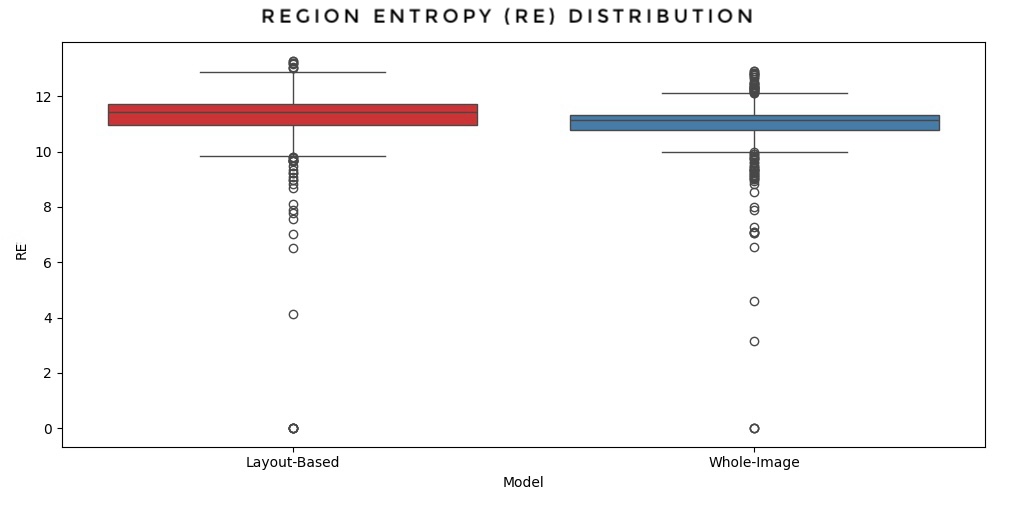}
 \caption{Distribution of Region Entropy (RE)}
        \label{fig:red_boxplot}
    \end{minipage}
\end{figure}

\section{Discussion}

This work aims to draw attention to the risk that, when AI document layout systems are developed without regard for cultural and historical context, the editorial logic and community priorities embedded in historical materials may be erased. Our early results show that using layout-aware OCR pipelines with these materials not only improves technical outcomes (reducing redundancy and increasing informational diversity), but also better reflects the creative and communicative practices that define these sources. This approach is consistent with responsible innovation in that it prioritizes cultural integrity and public trust, while also making clear the importance of participatory, justice-oriented AI development for historically marginalized collections.

The central goal remains to recover, transcribe, and make accessible the full breadth of Black press archives as accurately as possible. While layout-aware pipelines are an important step forward, highly irregular layouts and rare region types continue to pose significant challenges, and automated metrics cannot fully substitute for editorial expertise. Moving forward, achieving high-quality transcription will require close collaboration with Black historians, archivists, and community stakeholders to guide both evaluation and system development. Our next steps are to expand these methods and improve our tools with community feedback, and create AI systems (retrieval, generative, and beyond) that faithfully preserve the editorial logic and creative intent of Black newspapers. Through comparison, we also hope to provide critical analysis of existing systems and the ways they may fail to coherently address the unique spatial and linguistic meaning provided in historical Black press archives. Ultimately, our commitment to progress in this area will help us build an AI as a tool for honoring and preserving Black cultural memory.

\vspace{12pt}


\begin{thebibliography}{00}
\bibitem{b1} Ehrmann, Maud, Estelle Bunout, and Frédéric Clavert. "Digitised Historical Newspapers: A Changing Research Landscape." Newspapers—A New Eldorado for Historians (2023): 1-22.

\bibitem{b2} Lee, Benjamin Charles Germain, et al. "The newspaper navigator dataset: Extracting headlines and visual content from 16 million historic newspaper pages in chronicling america." Proceedings of the 29th ACM international conference on information \& knowledge management. 2020.

\bibitem{b3} B. Henry, ``The Representational Database: A Look at Black Newspapers and their Digitization History,'' National Endowment for the Humanities, 2024. [Online]. Available: https://www.neh.gov/blog/representational-database-look-black-newspapers-and-their-digitization-history

\bibitem{b4} The Crowley Company, ``Preserving Voices: Digitizing Howard University’s Historic Black Newspaper Collection,'' 2024. [Online]. Available: https://thecrowleycompany.com/preserving-voices-digitizing-howard-universitys-black-newspaper-collection/

\bibitem{b5} Howard University Moorland-Spingarn Research Center, ``Black Press Archive,'' 2025. [Online]. Available: https://msrc.howard.edu/black-press-archive

\bibitem{b6} Nieman Reports, ``The Black Press: Past and Present,'' 2024. [Online]. Available: https://niemanreports.org/the-black-press-past-and-present/

\bibitem{b7} Brigit Katz, ``The Archives of Historic Black Newspapers Are Going Digital,'' \emph{Smithsonian Magazine}, Aug. 2018. [Online]. Available: https://www.smithsonianmag.com/smart-news/explore-archives-historic-black-newspapers-180969292/

\bibitem{b8} Nazeem, Meharuniza, R. Anitha, and S. Navaneeth. "Open-Source OCR Libraries: A Comprehensive Study for Low Resource Language." Proceedings of the 21st International Conference on Natural Language Processing (ICON). 2024.

\bibitem{b9} Carlson, Jacob, Tom Bryan, and Melissa Dell. "Efficient ocr for building a diverse digital history." Proceedings of the 62nd Annual Meeting of the Association for Computational Linguistics (Volume 1: Long Papers). 2024.

\bibitem{b10} Li, Minghao, et al. "Docbank: A benchmark dataset for document layout analysis." arXiv preprint arXiv:2006.01038 (2020).

\bibitem{b11} Zhao, Zhiyuan, et al. "Doclayout-yolo: Enhancing document layout analysis through diverse synthetic data and global-to-local adaptive perception." arXiv preprint arXiv:2410.12628 (2024).

\bibitem{b12} Deng, Qilin, et al. "The YOLO model that still excels in document layout analysis." Signal, Image and Video Processing 18.2 (2024): 1539-1548.

\bibitem{b13} Zhong, Xu, Jianbin Tang, and Antonio Jimeno Yepes. "Publaynet: largest dataset ever for document layout analysis." 2019 International conference on document analysis and recognition (ICDAR). IEEE, 2019.

\bibitem{b14} Pfitzmann, Birgit, et al. "Doclaynet: A large human-annotated dataset for document-layout segmentation." Proceedings of the 28th ACM SIGKDD conference on knowledge discovery and data mining. 2022.
\bibitem{b15} Reul, Christian, et al. "Improving OCR accuracy on early printed books by combining pretraining, voting, and active learning." arXiv preprint arXiv:1802.10038 (2018).

\bibitem{b16} Kaltchenko, Alexei. "Entropy Heat-Mapping: Localizing GPT-Based OCR Errors with Sliding-Window Shannon Analysis." arXiv preprint arXiv:2505.00746 (2025). 
\bibitem{b17}Palacio-Niño, Julio-Omar, and Fernando Berzal. "Evaluation metrics for unsupervised learning algorithms." arXiv preprint arXiv:1905.05667 (2019).

\bibitem{b18} Ströbel, Phillip Benjamin, et al. "Evaluation of HTR models without ground truth material." arXiv preprint arXiv:2201.06170 (2022).
\bibitem{b19}Terven, Juan, et al. "Loss functions and metrics in deep learning." arXiv preprint arXiv:2307.02694 (2023).

\bibitem{b20} Terven, Juan, Diana-Margarita Córdova-Esparza, and Julio-Alejandro Romero-González. "A comprehensive review of yolo architectures in computer vision: From yolov1 to yolov8 and yolo-nas." Machine learning and knowledge extraction 5.4 (2023): 1680-1716.
\bibitem{b21} Gao, Zhangchi, and Shoubin Li. "YOLOLayout: Multi-Scale Cross Fusion Former for Document Layout Analysis." Available at SSRN 4423686 (2024).
\bibitem{b22} Saha, Ranajit, Ajoy Mondal, and C. V. Jawahar. "Graphical object detection in document images." 2019 International Conference on Document Analysis and Recognition (ICDAR). IEEE, 2019.
\bibitem{b23} Shehzadi, Tahira, Didier Stricker, and Muhammad Zeshan Afzal. "A hybrid approach for document layout analysis in document images." International Conference on Document Analysis and Recognition. Cham: Springer Nature Switzerland, 2024.
\bibitem{b24} Araújo, Alessandra Belézia. "Análise de layout de página em jornais históricos germano-brasileiros." 2019,


\bibitem{b25} Ströbel, Phillip Benjamin, and Simon Clematide. "Improving OCR of black letter in historical newspapers: the unreasonable effectiveness of HTR models on low-resolution images." (2019).

\bibitem{b27}Ströbel, Phillip Benjamin, et al. "Evaluation of HTR models without ground truth material." arXiv preprint arXiv:2201.06170 (2022). 

\bibitem{b28} Padilla, Rafael, Sergio L. Netto, and Eduardo AB Da Silva. "A survey on performance metrics for object-detection algorithms." 2020 international conference on systems, signals and image processing (IWSSIP). IEEE, 2020.

\bibitem{b29} Barlindhaug G. Artificial Intelligence and the Preservation of Historic Documents. Proceedings from the Document Academy (PDOCAM). 2022;9(2)
\bibitem{b31} National Endowment for the Humanities, “The representational database: A look at Black newspapers and their digitization history,” NEH Blog, 2024. [Online]. 
\bibitem{b32} Zytko, Douglas, et al. "Participatory design of AI systems: opportunities and challenges across diverse users, relationships, and application domains." CHI Conference on Human Factors in Computing Systems Extended Abstracts. 2022.

\bibitem{b33} Morrison, R. "Voluptuous Disintegration: A Future History of Black Computational Thought." DHQ: Digital Humanities Quarterly 16.3 (2022)
\end{thebibliography}
\end{document}